# A Trustworthy and well-organized data disseminating scheme for ad-hoc WSNs

Nomica Imran[1], Salman Khan[2], Imran Rao[3]

[1]Faculty of Information Technology, Monash University, Melbourne, Australia
`nmcho1@student.monash.edu.au`
[2]Department of Computer Engineering, Kyung Hee University, Korea
[2]`salman_r_a@yahoo.com`
[2]NICTA VICTORIA Labs, Department of Computer Science and Software Engineering
The University of Melbourne, Australia
[2]`imran@csse.unimelb.edu.au`

## ABSTRACT

*Wireless Sensor Networks (WSNs) generate massive amount of live data and events sensed through dispersedly deployed tiny sensors. This generated data needed to be disseminate to the sink with slight consumption of network resources. One of the ways to efficiently transmit this bulk data is gossiping. An important consideration in gossip-based dissemination protocols is to keep routing table up to date. Considering the inherent resource constrained nature of adhoc wireless sensor networks, we propose a gossip based protocol that consumes little resources. Our proposed scheme aims to keep the routing table size R as low as possible yet it ensures that the diameter is small too. We learned the performance of our proposed protocol through simulations. Results show that our proposed protocol attains major improvement in network reachability and connectivity.*

## 1. INTRODUCTION

A wireless sensor network (WSN) [17] is a multihop wireless network consisting of spatially distributed autonomous sensors with sensing, computation and wireless communication capabilities. Each sensor is generally a constrained device with relatively small memory, restricted computation capability, short range wireless transmitter-receiver and limited built-in battery. The deployment of sensors can be done in a random fashion (*e.g.,* airplane dropping in a disaster management scenario), or placed manually in strategic locations (*e.g.,* for intrusion detection or target tracking applications). These sensor nodes distributed over a given area are used to monitor and disseminate the collected events towards a base station, or sink, for post-analysis and processing. Typical sensed phenomena includes temperature, humidity, position, speed, motion, and others used in applications ranging from health care and logistics, through agriculture, forestry, civil and construction engineering, to surveillance and military applications.

Applications for wireless sensor networks [17] fall in three major categories: (i) *Periodic sensing:* sensors are always monitoring the physical environment and continuously reporting measurements to the sink, such as in weather monitoring applications; (ii) *Event-driven:* sensors operate in a silent monitoring state and are programmed to notify about events, such as the presence of objects in intrusion detection, target tracking or military applications; and (iii) *Query-based:* the generated data reports are kept available within the sensor network, and sensors react to the sink's queries by returning the corresponding requested measurements and events.

Since sensor nodes are energy limited and may fail at any moment, this data delivery is far from secure. Therefore, it is important to design novel solution to allow a robust and reliable data dissemination. Take the example of a geographical area being monitored for security breaches. Many dissemination protocols have been proposed to allow the dissemination of the collected events towards a static sink. All the events generated must reliably transmit to the sink node. Wireless ad-hoc networks are formed by a set of hosts that communicate with each other over a wireless channel. They provide an exclusive communication model. Each node has the ability to communicate directly with another node (or several of them) in its physical neighbourhood. They operate in a self-organized and decentralized manner and message communication takes place via multi-hop spreading. Any packet sent from one node to another may pass through a number of intermediate nodes acting as routers. The deployed ad-hoc WSNs pose great challenges in reliable sensed data delivery. Due to the small transmission range of sensor nodes the data is forwarded using multiple hops where unexpected node failure is common at each hop. Routing techniques in adhoc sensor network gives priority to reliable transmission as the loss of important information prevents these sensor networks from fulfilling its primary purpose and hence this information loss should be avoided.

The commonly used routing protocols uses single path routing or multiple path routing techniques to deliver the data, without providing reliable information dissemination and hence the overhead involve is same for all information. This paper presents a reliable data dissemination technique with efficient resource management for adhoc WSNs. The major contribution of the proposed scheme is to reduce the overhead of the acknowledgement. The proposed protocol is based on a simple idea of delaying the acknowledgement when and where possible. The acknowledgment is being delayed till it reaches the $N^{th}$ node conditionally the Nth node has the strength to send acknowledgement back to the source. At the same time, if the source node has the same signal strength, it will send the next data chunk directly to the $N^{th}$ node. Thus, saving considerable amount of network resources. Even the collision can be avoided by allocating the time slots plus ensuring the coordination among the nodes.

The proposed protocol works for ad-hoc WSNs and is not Applicable for WSNs because of relatively large number of nodes in the sensor networks. Again in WSN, it is not possible to build a global addressing scheme for the deployment of a large number of sensor nodes as the overhead of ID maintenance is high. Second, in contrast to typical communication networks, almost all applications of sensor networks require the flow of sensed data from multiple sources to a particular BS. This, however, does not prevent the flow of data to be in other forms (e.g., multicast or peer to peer). The remainder of the paper is organized as follows. Section 2 describes of the related work. Section 3 provides a detail description of our proposed protocol. In Section 4 we discuss the simulation results and analyse the performance and cost of our algorithm. Finally, we conclude the paper in Section 5.

## 2. RELATED WORK

Data dissemination protocols can be classified [17] according to several criteria. First, they vary in the nature of the disseminated information: *(i) data dissemination:* the measured data is disseminated; *(ii) meta-data dissemination:* a meta-data is disseminated while the measured data remains stored locally in the sensor; and *(iii) sink location dissemination:* the sink location is stored in the sensor field. When a node detects a new event, it determines the sink's location and the data is then forwarded to this location.

The protocols can also be classified depending on where the information is disseminated: *(i) a single node:* the disseminated information is stored in a particular node usually chosen in a deterministic and/or geographic way; *(ii) a node out of a group of nodes:* a group of nodes is defined and the information is disseminated towards one node out of this group, generally the closest to the source; and *(iii) a set of nodes:* the information is replicated over a set of nodes. Finally, protocols vary in the virtual infrastructure formed by the set of potential storing nodes. In general, virtual infrastructures can be divided into *rendezvous-based* approaches and *backbone-based* approaches. In both cases, the virtual infrastructure acts as a rendezvous region for the queries and the generated data.

The data dissemination techniques can further be categorized as structured and unstructured. The structured approach use up hash tables for table management routing. The same hash is used for placing data from different sources so as sinks uses to retrieve it. By doing so the query is significantly simplified as the sink knows where exactly should look for the stored data. The unstructured approach [1], [2], [3] implies the absence of a hash and the sink has no prior knowledge of the location of the information. In that scenario, the queries are disseminated to a randomly selected node. The surveys in [4] and [3] addressed several design issues and techniques for WSNs describing the physical constraints on sensor nodes, applications, architectural attributes, and the WSNS protocols proposed in all layers of the network stack.

Different approaches have been proposed in the literature for data dissimilation in ad-hoc WSNs. LAF (Location-Aided Flooding) [5] is based on a modified flooding. It makes use of location information to partition the network into virtual grids. Based on there location sensor nodes are grouped into a virtual grid. Nodes are categorized as gateway and internal nodes. The job of gateway nodes is to forward the packets across virtual grids where as internal nodes forward the packets within a virtual grid. Redundancy is being reduced in LAF by adding a special field in packet header called node list which contains the ids of all nodes already having the packet.

Flooding is not consider appropriate for WSNs even though it is merely a simple means of broadcasting [6] .The reason is flooding leads to collision and redundant packet reception that together deplete sensors of valuable battery power. Considering the load balance for conserving the energy of sensor nodes, multipath routing protocols which have the advantage of sharing energy depletion between all sensor nodes have been proposed [7]. However, no research has been conducted for the effects of route maintenance schemes on communication performance and energy efficiency.

Negotiation based protocols use high level data descriptors to eliminate redundant data transmissions through negotiation. Even the communication decisions are being taken depending on the available resources to them. The Sensor Network Protocols via Information Negotiation (SPIN) [8] and [9] are a set of protocols intended to disseminate data to all nodes in the network. The SPIN family of protocols uses data negotiation and resource-adaptive algorithms.

SPIN assign a high-level name to completely describe their collected data and perform meta-data negotiations before any data is transmitted. This assures that there is no redundant data sent throughout the network. The semantics of of the meta-data format is application-specific and is not specified in SPIN. We are considering the lossy broadcast medium protocols only [9]. Authors in [10] have pointed out that the SP1N-RL for a lossy broadcast network is not capable enough to ensure reliability. They argue that the performance of SPIN-RL suffers due to the lossy nature of the broadcast medium and is at the same time not capable enough to ensure information convergence at all nodes in the network. If a node misses an initial advertisement because of an undelivered packet means that it is un-aware of the availability of the data item and consequently cannot request it.

The scalable protocol for robust information dissemination, SPROID [10] is an energy-constrained, event-driven, reliable and efficient protocol. It recognizes the data generated by a unique tag .At the same time it uses the content tables for faster dissemination of information and guarantees reliable dissemination to all nodes in the network within a finite time. SPROID focus on the case of a single-channel broadcast medium. SPROID concentrate on a single-channel broadcast medium. SPROID achieves complete data dissemination in shorter time and with more energy efficiency as compared to SPIN [9].

Hue et al. [11] provides quick reliable dissemination of large data objects over a multi-hop, wireless sensor network. Each node infrequently advertises the most recent version of the data object to its neighbours. The node receiving an advertisement of older version will respond its object profile of new version. This process will go on until all the nodes get new version of data.

Directed diffusion is a novel data-centric, data dissemination technique. In Directed Diffusion [12] the data generated by the producer is named using attribute value pairs. The consumer node requests the data by periodically broadcasting an interest for the named data. Each node in the network will establish a gradient towards its neighbouring nodes from which it receives the interest. Once the producer detects an interest it will send exploratory packets towards the consumer, possibly along multiple paths. As soon as the consumer begins receiving exploratory packets from the producer it will reinforce one particular neighbour from whom it chooses to receive the rest of the data.

In acknowledgement-based reliable data dissemination protocols [16] data chunks are addressed and sent to only one receiver, which acknowledges each chunk packet received correctly. In doing so, a receiver should get the complete set of data chunks since chunks which have not been acknowledged are resent. But it doesn't address the issue of loosing the acknowledgement in between. Neither it overcomes the problem if any in-between node is corrupt or doesn't have enough energy to participate in data dissemination.

Membership protocol makes use of locally-maintained complete or partial list of all non-faulty members provided to each member that belongs to that group [13]. The protocol needs to make sure that any changes in membership wether because members joining, leaving or failing are made known to all non-faulty members of the group. Nearly all the membership protocols have two components: first one is to detect failures or node withdrawals and the second one is to spread the updates of membership information through the network. However, it is not possible for a failure detector to deterministically achieve both completeness and accuracy over an asynchronous unreliable network as shown by Chandra and Toueg in [14]. It resulted in the development of failure detection mechanisms that guaranteed completeness, although it achieves accuracy only in a probabilistic manner [15].

## 3. PROPOSED PROTOCOL

This unique technique ensures reliable data dissemination for unstructured ad-hoc WSNs in which the source node is not aware of the position of the sink node. The overall aim of the proposed scheme is to make sure that reliability and efficiency is not being compromised. N is the reliability factor. The underline idea is to delay the acknowledgment until it reaches the $N^{th}$ node. Now, to ensure the reliable reception of data, the $N^{th}$ node will acknowledge the data reception message to the source directly. We assume the value of $N^{th}$ node bounded by network signal strength and is being set by the network administrator.

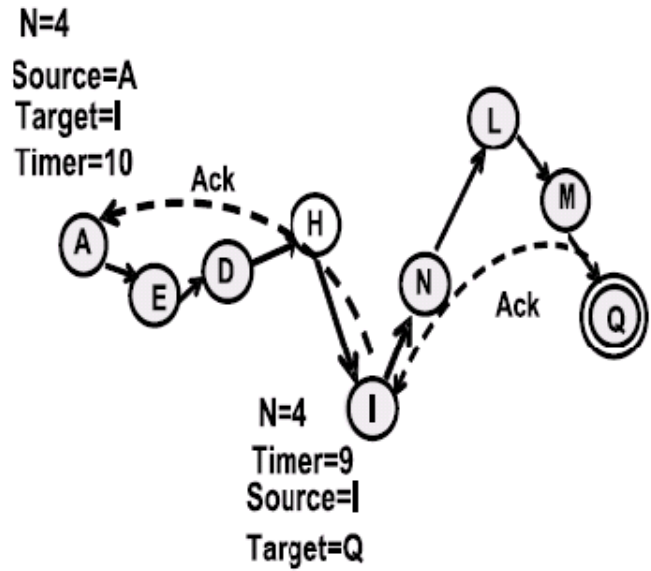

Figure 1. Overview Of the Proposed Scheme

Network traffic can be reduced by increasing the value of N. Furthermore, our proposed protocol ensures high fault tolerance by avoiding faulty sensors and in turn it increase the network life time. The overview of the proposed scheme is presented in Fig. 1. In the rest of this section we explain in details the working of our proposed membership management protocol along with the message routing scheme.

**An Example**

*A* wants to transfer data to the destination node. In this case *Q* is the destination node. *A* doesn't know the location of the node. *A* will set the value of $N^{th}$ node as 4. It means that *A* will traverse four neighbours. In fig, *A* will send data to *E, D*, and *H* and than reaches to $I^{th}$ node. The timer is set to 10 seconds. As the value of $n^{th}$ node is expired at *I*, $I^{th}$ node has received the data and after receiving the data chunk will send the acknowledgement back to *A* ensuring that it has received the data in defined time slot. Now *I* has become the source node. Again the value of $n^{th}$ node has been set along with the timer. The timer has been set to 9 sec and the $n^{th}$ value to 4. By doing so the data will reach to $Q^{th}$ node which is the destination node.

Following issues are observed in the scheme. We will start by considering the first scenario in which an in-between node has not received the acknowledgement in time. If an in between node has not received the acknowledgement in time it means that either the timer has expired or an in-between node is corrupt. We will see both of these cases. Starting with the first case in which the timer has expired earlier before reaching the node, as shown in fig 2. If the timer has expired and acknowledgement has not received in time, merely resetting the value of timer will help. This problem can be fixed by simply changing the value of timer from 10 to 15 sec will fix the problem.

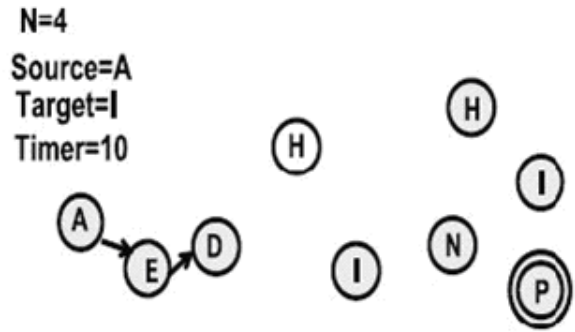

Figure 2. Missed Acknowledgment

Now we will consider the second case in which the in between node is corrupt. This case is explained in detail in fig 4. As shown in fig, the n$^{th}$ value is set to 4 and the timer is being set to 10 sec. But *A* didn't get the acknowledgement back in time. Even the resetting of the timer doesn't work. The process is being repeated by changing the value of n$^{th}$ node as 2. Even after that *A* didn't get the acknowledgement back. Again the process is being repeated by setting the value of n$^{th}$ as 1. *A* didn't get the acknowledgement back. It means that *E$^{th}$* node is corrupt. This problem can be overcome by rerouting the packet with a different route.

Next we will consider the third scenario in which *I* don't have enough energy to send the data. *I$^{th}$* node will check the status of its energy level. If it is below a specified threshold, the node will not participate in the process. If it has energy but limited in amount, it will try to send data to its neighbour node.

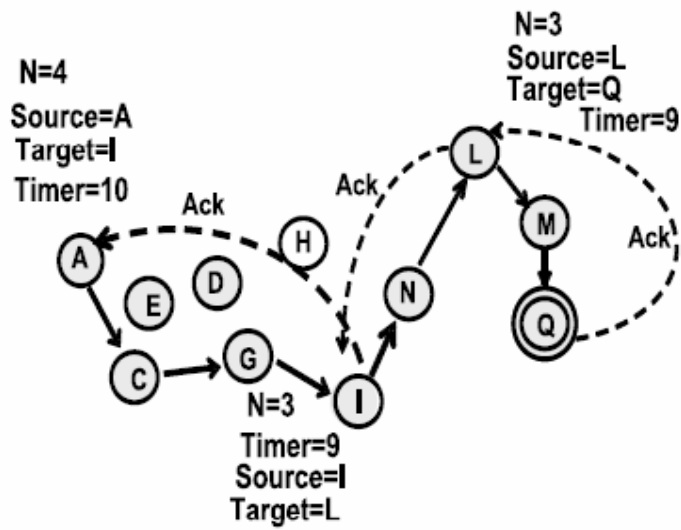

Figure 3. Weak In-Between Node (Option 1)

In the fig, *I$^{th}$* node sends the data to *k$^{th}$* node in its immediate neighbourhood by keeping the value of n to be 1. I will in the meanwhile send an acknowledgement back to *A* showing its energy level and at the same time *A* will get an acknowledgement from k .Now *A* knows that the data is with *K$^{th}$* node. *K$^{th}$* node will also send an acknowledgement back to *I$^{th}$* node to confirm that it has sent the acknowledgment back to *A*. Now *K* has become the source node.

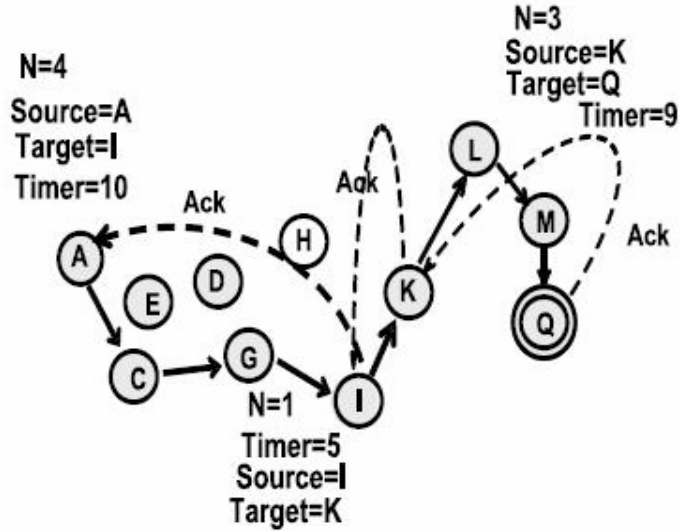

Figure 4. Weak In-Between Node (Option II)

Let there are N nodes in the network. Every node **p** keeps a routing table *Tp* of *I* arbitrary nodes in its routing table.

```
1  PROCEDURE: nodeJoin(node u)
2  begin
3      Node bootstrapNode := b
4      Int maxCount := I
5      Vector bList := b.receiveList(maxCount)
6      for each peer p ∈ bList
7      begin
8          updateRTable(u, tuple(p, l_p))
9      end
10     updateRTable(u, tuple(u, l_u))
11 end
```

Figure 5. Join Alogorithm

```
1  PROCEDURE: nodeMembership(node u)
2  begin
3      Node tmp := getRandomNode()
4      Vector bList := tmp.getList(I)
5      for each node a ∈ bList
6      begin
7          u.updateRTable(a)
8      end
9  end
```

Figure 6. Node Membership Algorithm

When node p first time join the network (or reconnect to the system after a (un)intentional departure), it executed the Join Algorithm as shown in Fig. [5]. This list of neighbours is maintained through a periodic refresh algorithm as shown in Fig. [6].

### 3.1. Message routing Algorithm

The system invokes Algo. when a node *p*, interested in to send data to another node, says the node *q*. The node *p* will dispatch message to its randomly selected neighbour *u*. The node u upon request to forward data *d* to node *q*, store the *d* in its local cache and initialize the *K* counter. The data chunk *d* is kept there until acknowledge for it is received. The node *u* forwards the data to the first node w in *Tu*. Node *u* waits for the acknowledgement from' the $K^{th}$ node. If the acknowledgement is not received until an appropriate time *t, u* assumes that w is not available and forwards *d* to the next node in *Tu*. When adjusting value of *t*, it is important to note that *t* is directly proportional to *K*, more the value of *K*, more u has to wait for the acknowledgement to arrive.

```
1  PROCEDURE: startRouting(this-node p, destination q, data d)
2  begin
3  |    routeMessage(p, p, q, d)
4  end
```

```
1  PROCEDURE: sendMessage(this-node w, source u, destination q, data
     d, K)
2  begin
3  |    IF(q == w)
4  |    begin
5  |    |    sendAck(u)
6  |    |    EXIT()
7  |    end
8  |    K := K - 1
9  |    Node tmp := random node ∈ T_w
10 |    IF( K ≥ 0)
11 |    begin
12 |    |    sendMessage(tmp, u, q, d, K)
13 |    end
14 |    ELSE
15 |    begin
16 |    |    sendAck(u)
17 |    |    routeMessage(tmp, w, q, d)
18 |    end
19 end
```

Figure 7. SendMessage Algorithm

```
1  PROCEDURE: routeMessage(this-node w, source u, destination q, data
   d)
2  begin
3      DEFINE K := Reset K
4      DEFINE t := time to wait for ack
5      DO: Store d in local cache
6      DO-LOOP
7      begin
8          Node tmp := next node ∈ T_w
9          u.sendMesage(tmp, u, q, d, K)
10         /* loop until ack received OR t expired */
11         DO-LOOP
12         begin
13             IF ( ack-received )
14             begin
15                 ack-received := TRUE
16                 break
17             end
18             t := t - 1
19         end
20         UNTIL ( t ≥ 0 )
21         /* if ack is received exit the function*/
22         IF ( ack-received == TRUE )
23         begin
24             DO: remove d from cache of u
25             EXIT()
26         end
27         /* otherwise remove tmp and continue through next node in
           T_w*/
28         ELSE
29         begin
30             DO: Remove tmp from T_w
31             DO: Continue the loop
32         end
33     end
34     UNTIL (T_w is not empty)
35 end
```

Figure 8. Route Message Algorithm

Node w decrease the **K** counter and forwards the chunk to the first node in its routing table **Tw**. If **K** is not equal to zero, node **w** will pick one node at random from its routing table **Tw** and forward **d** to it. The d is routed in the network until it reaches its destination or K becomes zero. If the value of **K** is equal to zero, an acknowledgement is sent back to the source and the node w will declare itself as the new source of the data **d** and execute Route Message.

**3.3. Discussion**

The major contribution of the proposed scheme is to reduce the overhead of the acknowledgement. The proposed protocol is based on a simple idea of delaying the acknowledgement when and where possible. The acknowledgment is being delayed till it reaches the $N^{th}$ node conditionally the $N^{th}$ node has the strength to send acknowledgement back to the source. At the same time, if the source node has the same signal strength, it will send the next data chunk directly to the $N^{th}$ node. Thus, saving considerable amount of network resources. tr

We can also find out the sink location and at the same time calculate the network reliability by disseminating test packets. On receiving the acknowledgement, we can further increase the value of N to make the scheme more efficient and can reduce the network traffic. At the same time, if the acknowledgement is not received in dedicated time, we can decrease the value of N to ensure that the acknowledgement is not missed because of weak signal strength of $N^{th}$ node. It possibly will be argued that if the $N^{th}$ node is within the source range, the performance can be enhanced by sending the data packet directly to the acknowledging node without relying on the intermediate nodes. At this juncture we should clear the point that in the absence of complete list of other nodes, there can be a possibility that the source node may not know about the exact location of the destination node. Again, if the destination node is within the signal strength of sending node, it will next time send the data directly to it. By doing so, we ensure secure data delivery yet by keeping the resources low. Even the collision can be avoided by allocating the time slots plus ensuring the coordination among the nodes.

## 4. SIMULATION RESULTS

We simulated our proposed algorithm for 10 different values of K. We injected the faults into our system at different fixed rates and analysed the communication cost. System was tested for a data dissemination spawning 100 hops. The communication cost calculated as the number of bytes to disseminate a message from source to destination and the system efficiency is calculated as ratio of the time taken to send data from source to destination with faults and without faults. If the target node is within the transmission range of the data source (i.e. the hop to reach sink node are less then or equal to K), no further overhead is required.

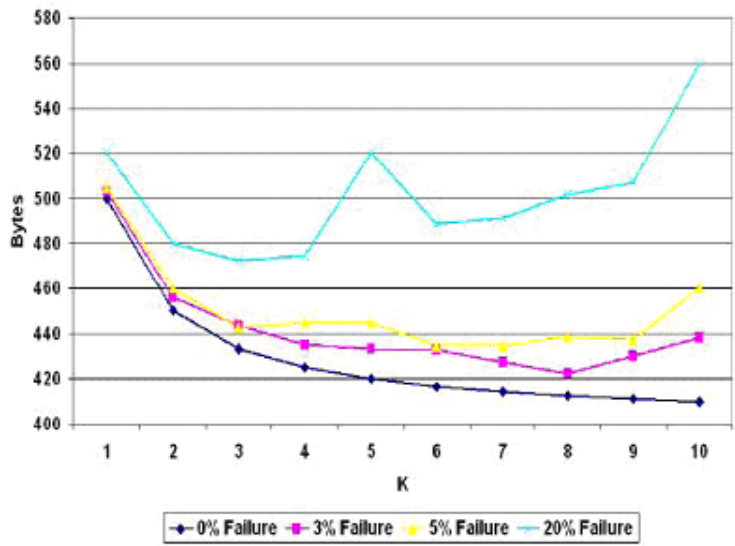

Figure 9.  Communication Cost for varying values of K and varying faults rate.

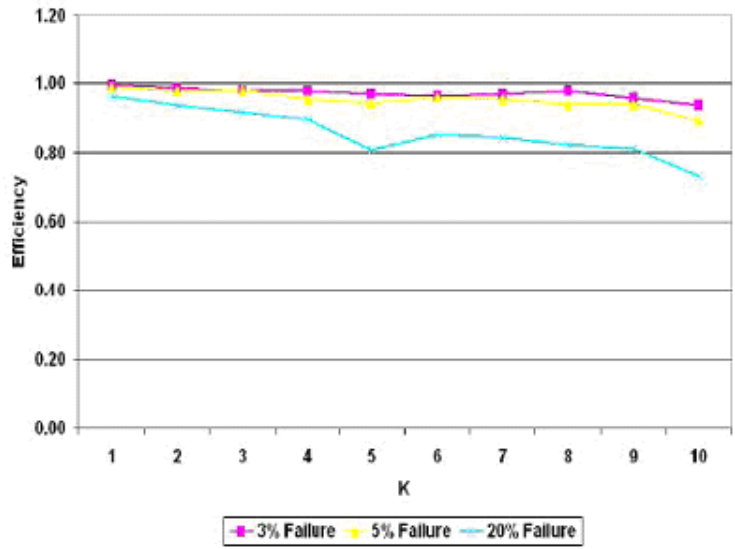

Figure 10. Efficiency of the proposed protocol in case of varying faults.

However, in order to account for multi-hop networks, the intermediate nodes will act as a data sources in order to cover distant nodes.

Decreasing the value of K will give more reliability (that is an acknowledgement after every hop). Another advantage of keeping K low is that if message is lost very close to destination node, data chunk is sent to the next node from the new source. This, hence, save us from routing the data chunk again all the way back from the original source p. Moreover, the data is routed through a new node hence increasing the chances of successful delivery. On the other hand, for comparatively stable network links, increasing the value of K will result in high efficiency of the routing protocol as shown in Fig. 10.

As seen in Fig. 9, our proposed scheme ensures the reliability against odd failures where acknowledgement cost can be mitigated or reduced without compromising on the reliability of the system. For highly un-reliable networks, the cost of acknowledgement is very less than the actual data dissemination and thus the benefit of our scheme cannot be realized.

## 5. CONCLUSION

In this paper, we present a reliable and efficient data dissemination scheme for our proposed protocol. The proposed protocol considers the wireless lossy channel confronted by sensor networks. The basic idea we employ is to delay the acknowledgement message until it reaches $N^{th}$ hop. The $N^{th}$ node than acknowledge the receipt of the message to the source directly. By increasing the value of N, the network traffic can be minimized. Our proposed protocol is adaptive and self-configurable to churn transient failures. Moreover the size of routing table is decreased resulting in increase in efficiency.